# Spectro Capture: A software system for automated small-observatory spectroscopy

Paul Luckas *

*Shenton Park Observatory, Perth, Western Australia, Australia*
*International Centre for Radio Astronomy Research, The University of Western Australia, 35 Stirling Hwy, Crawley, Western Australia, Australia*

ABSTRACT

Spectro Capture is a Python-based software system developed to automate small-observatory fibre-fed spectroscopy. The system integrates target selection, telescope slewing, guide star acquisition, fibre position restoration, calibration and science exposure sequencing within a single observing workflow. The paper describes the design and operational behaviour of the system at Shenton Park Observatory, where it has been used for unattended multi-target spectroscopic observing. Log analysis from January to May 2026 shows that the system completed 339 of 345 attempted science target blocks in primary unattended batch runs, corresponding to a completion rate of 98.3%. The results demonstrate that reliable unattended spectroscopy is practical at a small observatory when target acquisition, guiding, scheduling and calibration control are treated as an integrated software problem.

## 1. Introduction

Automation has been part of observational astronomy for many years. Robotic telescope systems routinely schedule targets, respond to weather conditions, control domes, acquire images and process data with little or no human intervention. For imaging and survey work in particular, unattended operation has become routine at both professional and amateur observatories.

However, spectroscopic observing has proven more difficult to automate. While exposure sequencing, calibration management and scheduling are readily handled using standard scripts and techniques, automatically acquiring a target and keeping it correctly positioned on a spectrograph entrance remains problematic. As a result, many spectroscopic systems, especially at smaller observatories, still depend on human intervention during target acquisition and guiding.

The difficulty is largely practical. In spectroscopy the target must be placed on a very small entrance aperture, whether a slit or a fibre, often only tens of microns across. In many imaging systems this corresponds to only a few guide camera pixels. Once positioned, the target must be held there for the duration of an exposure that may last tens of minutes or longer while seeing, transparency, telescope tracking and guiding performance continue to vary. Guider fields of view are typically small and may offer few suitable guide stars beyond the target itself. Even when surrounding stars are present, plate solving and mount repositioning are often unable to deliver the sub-pixel accuracy needed to centre a target reliably on the spectrograph entrance.

Guiding presents further complications. Because the field of view is often limited, guiding is commonly performed on the target itself. In this situation much of the target's light is directed into the spectrograph rather than the guider, reducing the available guiding signal and making accurate guide star tracking difficult. As a consequence, guiding performance has a direct and immediate effect on the quality of the spectroscopic data.

Traditionally, tasks associated with target positioning and guiding have been handled by the observer. Target placement is confirmed visually, guider parameters are adjusted as conditions change, and intervention occurs when performance becomes marginal. Much of this process depends on experience rather than explicit rules, which has made it difficult to translate into reliable automated behaviour.

This paper describes an attempt to address these problems with a software application developed by the author to support fully automated spectroscopic observing. The aim is not to show that software can outperform an experienced observer during a single manually supervised target acquisition, but to improve the practical efficiency of the observatory by allowing multi-target spectroscopy to proceed unattended over the course of a night. In this context, efficiency means reducing the need for repeated observer intervention in the tasks required to acquire spectra, rather than demonstrating that manual observing is slower. The discussion is grounded in operational use at a small suburban observatory under varying sky conditions.

* Correspondence to: International Centre for Radio Astronomy Research, The University of Western Australia, 35 Stirling Hwy, Crawley, Western Australia, Australia.
*E-mail address:* paul.luckas@uwa.edu.au.

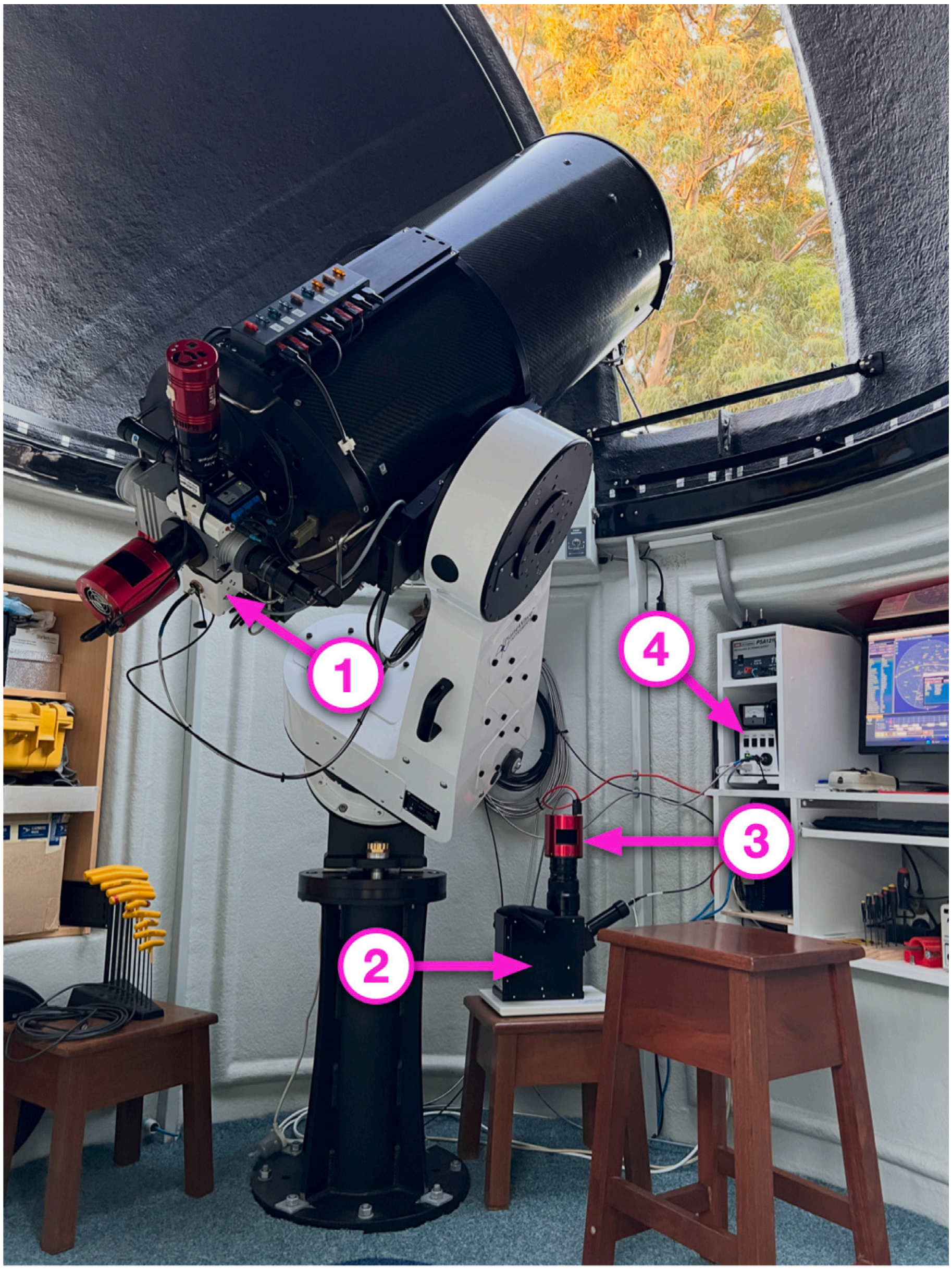

**Fig. 1.** Labelled overview of the Shenton Park Observatory spectroscopy installation. Numbered components are: (1) fibre injection and guiding unit, with fibre optic cables routed down and through the mount; (2) Shelyak eShel spectrograph; (3) spectrograph acquisition camera; and (4) thorium and tungsten lamp control unit with supporting power/control hardware. The telescope is mounted on a PlaneWave L-350 direct drive mount inside the observatory dome.

## 2. Observatory configuration and baseline workflow

### 2.1. Instrumentation

The software described in this paper was developed and tested at Shenton Park Observatory.[1] The observatory is located only a few kilometres from the centre of the city of Perth in Western Australia, and is affected by significant light and atmospheric pollution. Despite these challenges, the climate is well suited to long periods of clear skies during the summer months which has allowed for almost continuous observing during the period of testing. The instruments include a 0.35 m RC14C Ritchey–Chrétien telescope (DeepSky Instruments, 2026) mounted equatorially on a PlaneWave L-350 mount (PlaneWave Instruments, 2026) and housed in a motorised 2.3 m observatory dome.

The PlaneWave L-350 mount is operated using PlaneWave Instruments' PWI4 control system, which provides network-based control and status monitoring. The current pointing model was created on December 6, 2025 and contains 127 all sky points down to an altitude of 33 degrees. The reported RMS residual is 12.7 arcseconds. Both mount and dome are computer interfaced and are capable of automated operation under software control.

[1] Shenton Park Observatory is a private observatory located in Perth, Western Australia. It is operated by the author and is not a public facility.

The application described in this paper was developed specifically for operation with a Shelyak eShel spectrograph (Shelyak Instruments, 2026) which is located adjacent to the telescope inside the observatory. Light from the telescope is injected via the Shelyak fibre injection and guiding unit (FIGU) using a 50 μm optical fibre. A second fibre path is used to inject tungsten and thorium calibration light separately into the FIGU, allowing flat field and wavelength calibration frames to be acquired without changing the fibre connection. The eShel employs a 79 lines $mm^{-1}$ reflection grating in combination with a prism cross disperser, providing simultaneous multi-order wavelength coverage from approximately 3850 Å to 7400 Å in a single exposure. Spectral images are recorded using a ZWO ASI2600 series CMOS camera (ZWO, 2026) mounted at the spectrograph output and coupled via a Canon FD 135 mm f/2 photographic lens used as the camera optic. The ASI2600 detector employs a 26-megapixel APS-C format sensor with 3.76 μm square pixels and is operated in 2 × 2 on-chip binning mode for all observations. Under normal operating conditions the instrument delivers a spectral resolving power of $R \approx 13{,}000$.

Target acquisition and guiding are performed using a dedicated ZWO ASI174 mini guide camera attached to the FIGU. The guider views the telescope focal plane directly and provides continuous closed-loop guiding during all spectroscopic exposures. The usable guiding field of view is approximately 10 arcminutes × 6 arcminutes owing to the limited optical field delivered by the FIGU which is further reduced by significant geometric distortion towards the edges of the field. As a consequence, accurate target acquisition and guiding are restricted to a relatively small central region of the guider field.

The telescope mount, optics, FIGU, and guide camera form a rigid, permanently aligned assembly, permitting rapid target acquisition and highly repeatable positioning between observing sessions. This configuration provides a stable and efficient platform for automated stellar spectroscopy under the light polluted suburban skies of Perth, Western Australia.

### *2.2. Previous software environment and workflow*

Prior to the development of the software described here, routine ‘single target’ spectroscopic observing was carried out using a conventional ‘imaging-centred’ software stack comprised of applications such as MaxIM DL (Diffraction Limited, 2026b) for image acquisition, and CCD Commander (CCD Commander, 2026) for sequencing. Several guiding software solutions have been used over the years, with PHD2 (PHD2 Guiding, 2026) the preferred application.

The level of automation employed with this system was limited to single target work using software not designed with spectroscopy as a primary use case. Sequencing logic, calibration handling, and feedback mechanisms largely reflect a ‘conventional imaging’ workflow, with target acquisition, positioning and spectral image checks all performed manually.

These limitations, together with declining support for the existing software tools such as CCD Commander, motivated the development of a dedicated spectroscopic control platform.

## 3. Design and evolution of the software solution

### *Python framework*

Early on it was decided that the software should be developed as a modular Python (Python Software Foundation, 2026) application with an integrated graphical interface organised around the main observing tasks.

The framework provides the following core functionality:

- direct camera control via ASCOM (ASCOM Initiative, 2026), including exposure control, binning configuration and temperature management
- telescope and dome connectivity for control, monitoring and synchronised operation
- PHD2 guide integration for target acquisition, positioning and guiding
- deterministic creation of calibration and science data directories and file names
- execution of calibration and science exposure sequences, including calibration lamp control
- FITS (Wells et al., 1981) file generation and header population

These items form the foundation upon which the automated acquisition and guiding system has been built. Together they provide the basic cycle required for unattended spectroscopy:

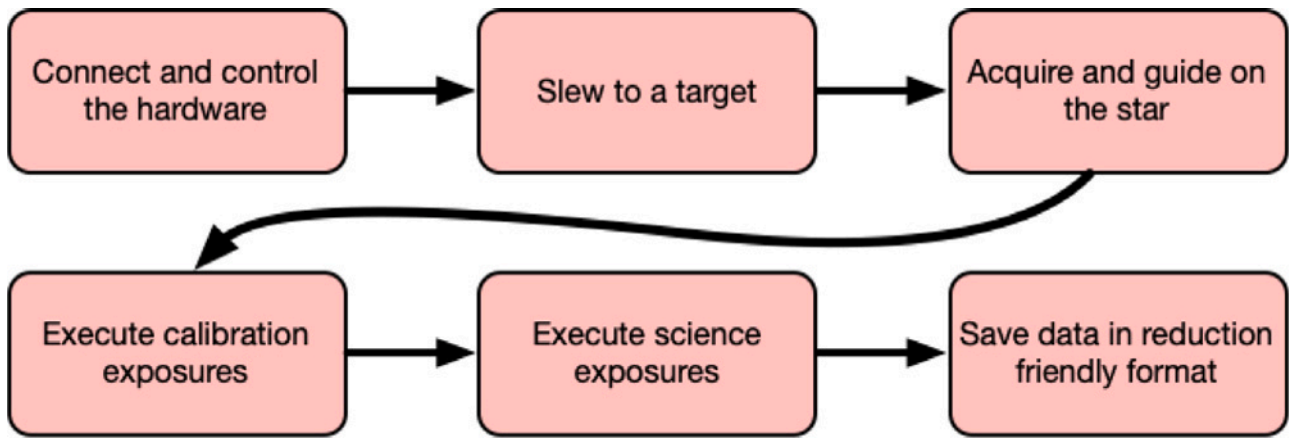

The software, referred to from this point onward as *Spectro Capture*, also includes several supporting features that are useful in routine operation but are not central to the automation framework. These include:

- target database tools
- SIMBAD (SIMBAD Astronomical Database, 2026) lookup support
- manual dark and bias frame acquisition
- manual calibration lamp control
- a minimal integrated FITS viewer for inspecting live spectral images

The application’s user interface provides access to controls and features through tabs corresponding to individual Python scripts. A lightweight shared application context is used to coordinate state, configuration and device connections across these scripts.

## 4. Development pathway

### *4.1. Auto capture*

The framework described above evolved into its present form through a sequence of increasingly automated workflows, some of which remain visible in the current application as legacy modes of operation. Development began with the most difficult practical problem: acquiring the target star and placing it accurately on the fibre. In this early mode, the telescope was slewed manually so that the target appeared ‘somewhere’ within the guide camera field. The software then acquired the star, started guiding and moved it to the stored fibre position. Calibration and science exposures were configured and executed separately.

This acquisition-and-guiding workflow was refined over several months of live testing. Its purpose was to establish that repeatable guide acquisition and fibre positioning could be handled under software control without user intervention.

The *Auto Capture* script extended this approach by combining target acquisition and guiding with telescope slewing and exposure sequencing, demonstrating that the full acquisition-to-exposure cycle could be automated. However, Auto Capture remained essentially a single-target workflow and has since been superseded by the *Batch Runner* script as the main focus of ongoing development.

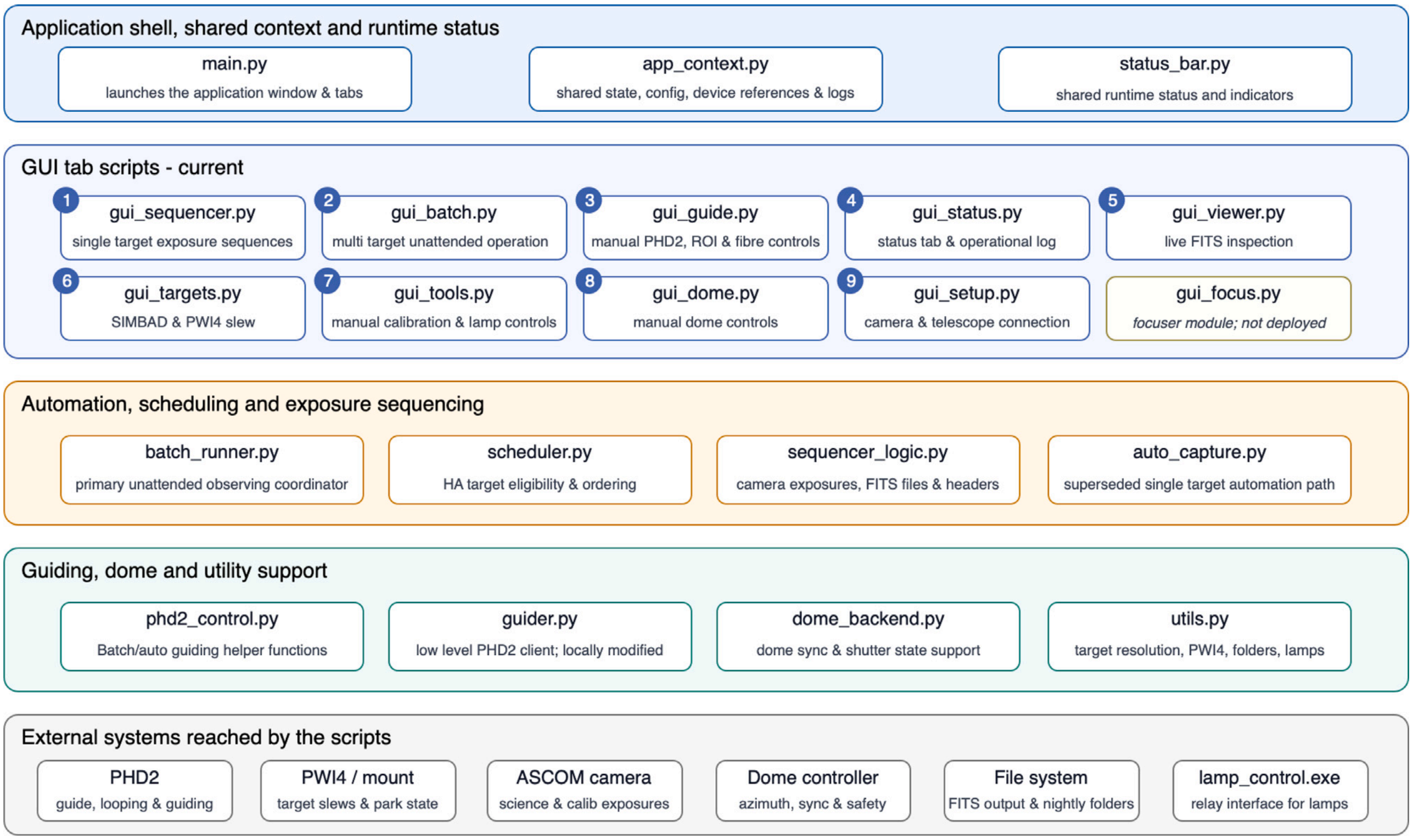

**Fig. 2.** Script architecture of the Spectro Capture application.

### 4.2. Batch Runner

Rather than treating each target as a separate operation, Batch Runner accepts a list of targets and associated observing parameters and then works through them sequentially using the same underlying acquisition, guiding and exposure pipeline developed earlier.

This shift is important because the practical value of automated acquisition and guiding is not simply that one target can be observed without intervention, but that a sequence of targets can be planned and acquired over the course of a night. The Batch tab has therefore become the main working interface in Spectro Capture. Target lists can be prepared in advance and the software then works through the observing plan completely unattended.

### 4.3. Automatic scheduling

An additional scheduler component adds a further layer to the batch process by determining when individual targets in a large list are permitted to start based on observing constraints rather than fixed timing — notably an 'hour angle'. The scheduler is responsible only for target eligibility and ordering; acquisition, guiding, calibration and exposure control are still delegated to the common automation scripts and functions. The detailed behaviour of Batch Runner and Scheduler and their interaction with the acquisition and guiding system, are described in later sections.

## 5. System architecture and core modules

The entire application is implemented as a set of Python scripts and dependencies grouped by function rather than as a single monolithic program. The graphical interface is implemented using Python's built-in `tkinter/ttk` GUI toolkit and is launched from `main.py`. It presents as a set of window tabs corresponding to the main observing tasks: `gui_batch.py`, `gui_guide.py`, `gui_setup.py` and so on. Operational behaviour is then handled by lower-level control scripts, including the Batch Runner, guiding, sequencing, dome and utility modules. Shared state, device connections and logging are coordinated through `app_context.py`.

Fig. 2 details the main script structure of Spectro Capture.

Although included in the diagram, descriptions of specific control software such as the mount and dome control software, ASCOM, PHD2 and eShel lamp module control are considered outside the objectives of this paper.

### 5.1. Guiding system logic

The guiding logic comprises four main steps:

1. Confining star detection to a defined region of interest.
2. Using a controlled exposure sequence to find the target.
3. Restoring the detected star to a stored fibre reference position.
4. Using adaptive exposure control once guiding is established.

Acquisition and guiding logic are built on PHD2's external control interface and the Python PHD2 client script developed by Galasso (2026). In this implementation, PHD2 provides the low-level guiding control, while Spectro Capture's custom Python scripts coordinate star selection, exposure adjustment and target-on-fibre restoration.

After the telescope has slewed to the requested target and settled, PHD2 is placed into looping mode. The software then searches for a star within a defined rectangular region of interest (ROI) in the guide camera image. This ROI is specified by a configurable centre coordinate and $x$ and $y$ dimensions and is positioned around the area where the target is expected to appear after a normal telescope slew.[2] It does not necessarily correspond to the fibre position. Restricting the search to this area helps PHD2 select the intended target rather than a brighter unrelated field star.

Target finding is handled by the ROI acquisition function in `batch_runner.py`. It uses an exposure ladder beginning with a 0.01 s guide exposure and progresses through longer exposures until PHD2 detects a star within the configured ROI. The actual PHD2 star selection request is issued through the modified `guider.py` client, which passes the ROI rectangle to PHD2's `find_star` command.

Once a suitable star has been detected, PHD2 guiding is started on that star and the adaptive exposure thread begins. Initially this operates in a 'pre-fibre' mode, where the aim is to maintain a workable guide signal while the star is being moved to the fibre position. The exposure logic uses a guide star brightness proxy derived from PHD2's star mass and HFD measurements,[3] with broad limits of 2000 to 65,000 PHD2 star mass counts during this phase so that it can respond quickly to weak or bright guide signals. The general aim is to move the target onto the fibre as efficiently as possible.

The fibre position is determined during setup and entered as the *Fibre Lock Position* on the Guide tab. The software sends this saved coordinate to PHD2 using its lock-position mechanism, so that the fibre position becomes the guiding reference rather than the position where the star was first detected. This distinction is important: the ROI centre, detected star position and fibre position are separate coordinates.

Successful centring is verified rather than assumed. The script first checks that the star has reached the broader fibre region, then applies a stricter stability test before science exposures are allowed to begin. This logic requires the target to remain close to the fibre for several consecutive guide-camera updates, using configurable tolerances and stability values that can be adjusted for the guide camera pixel scale and observing conditions. In practical terms, this prevents the exposure sequence from starting while the star is still settling, or while guide correction is responding to a transient measurement. Conversely, the timing and stability values can also be relaxed for less than optimal conditions.

After centring has been confirmed, the adaptive exposure thread switches into 'fibre mode'. In this mode the guide signal is held within a narrower target range of 25,000 to 55,000 PHD2 star mass counts by stepping the guider exposure up or down through configured exposure presets. This deliberately favours longer guide exposures, reducing the tendency for the guider to chase short timescale seeing fluctuations. Science exposures then proceed with the star verified on the fibre and guiding stabilised.

The time it takes to move from PHD2 looping to the start of science imaging is typically less than a minute, depending mainly on the distance between the ROI centre and the fibre. It also depends on target brightness, since fainter targets require longer guide exposures during the restore phase.

Fig. 3 summarises the full acquisition and guiding sequence.

### 5.2. Telescope and dome coordination

In the present version, telescope slews are handled by the central `slew_target()` routine in `utils.py`, which sends commands to the PlaneWave PWI4 control system through its HTTP interface rather than through an ASCOM telescope slew command. This provides a common slewing pathway across the application and avoids inconsistent coordinate handling in the tested PWI4-based implementation.

The dome is controlled separately from the telescope but is included in the same target acquisition sequence. Dome synchronisation is handled by `dome_backend.py`, which follows the telescope position and moves the dome when required. During an active telescope slew, normal dome-sync updates are paused and resume after the telescope has completed its slew.

Before guiding begins, `batch_runner.py` calls a settle routine that requires two consecutive stationary readings for both telescope and dome. This prevents PHD2 acquisition from starting at the instant a motion flag clears, giving both the mount and dome time to stabilise. 'Closed' or 'error' shutter states are checked at several points during the batch process, and repeated unsafe readings cause the batch to abort, request a PWI4 park and set the batch stop flag. This provides a basic 'weather aware' safeguard for an observatory fitted with a cloud sensor.

Such a 'safeguard' was added midway through the observing period with the integration of a Boltwood Cloud Sensor (Diffraction Limited, 2026a). The sensor hardware interfaces with the dome shutter controller and provides weather state information including the onset of cloud or rain. In the current implementation, an unsafe weather state simply triggers a dome shutter closure which then actuates the aforementioned interrupt condition in `batch_runner.py`. The present paper does not treat weather sensing or weather recovery as a primary performance result.

This telescope/dome coordination layer is important because the guiding system assumes that the target will appear inside a relatively small, predefined ROI. The telescope, dome and settle logic therefore form the mechanical hand-off between target scheduling and the guide-acquisition sequence described earlier.

### 5.3. Camera control and FITS generation

Science camera control is provided through the ASCOM camera driver selected in the Setup tab. Exposure and file writing logic is handled by `sequencer_logic.py`, principally in the `expose_and_save()` function. This routine is used by both manual Sequencer operations and automated Batch Runner operation, so that calibration and science frames are acquired through the same camera control pathway.

For each exposure, `expose_and_save()` sets the configured binning, resets the camera to a full-frame readout, starts the ASCOM exposure, and then polls the camera's *ImageReady* state until the frame is available. A timeout margin is included so that the application does not wait indefinitely if the camera fails to report completion. During longer exposures the routine also updates the status display once per second, providing visibility of the current image number and elapsed exposure time.

When the camera reports that the image is ready, the image array is downloaded from the ASCOM driver and checked against the expected binned frame dimensions. During development, the ASCOM `ImageArray` was found to be returned with its axes reversed relative to the expected binned camera dimensions. The routine therefore transposes the array before saving so that the FITS files have the expected orientation for later reduction.

The data are then normalised into the unsigned 16-bit range and written as signed 16-bit FITS data using BZERO = 32768 and BSCALE = 1. This preserves compatibility with the FITS handling used by common spectroscopy reduction software.

FITS header creation is handled by the companion `fill_fits_header()` routine in `sequencer_logic.py`. This records the exposure timing, image type, object name, binning, observatory information, camera temperature and relevant telescope metadata. A telescope connection through the application context is used to populate relevant telescope related FITS header fields, even though target slews are issued to PWI4 directly through its HTTP interface.

[2] Although greatly dependent on mount quality and pointing stability, a configurable ROI allows less than perfect mount performance to produce surprisingly repeatable results, particularly with bright targets.

[3] PHD2 star mass is an empirical integrated brightness measure reported by PHD2 for the detected guide star. It is derived from the guide star signal relative to the local background and is used here only as a guide camera count based proxy for exposure control. It should not be interpreted as a calibrated flux or as a peak ADU value.

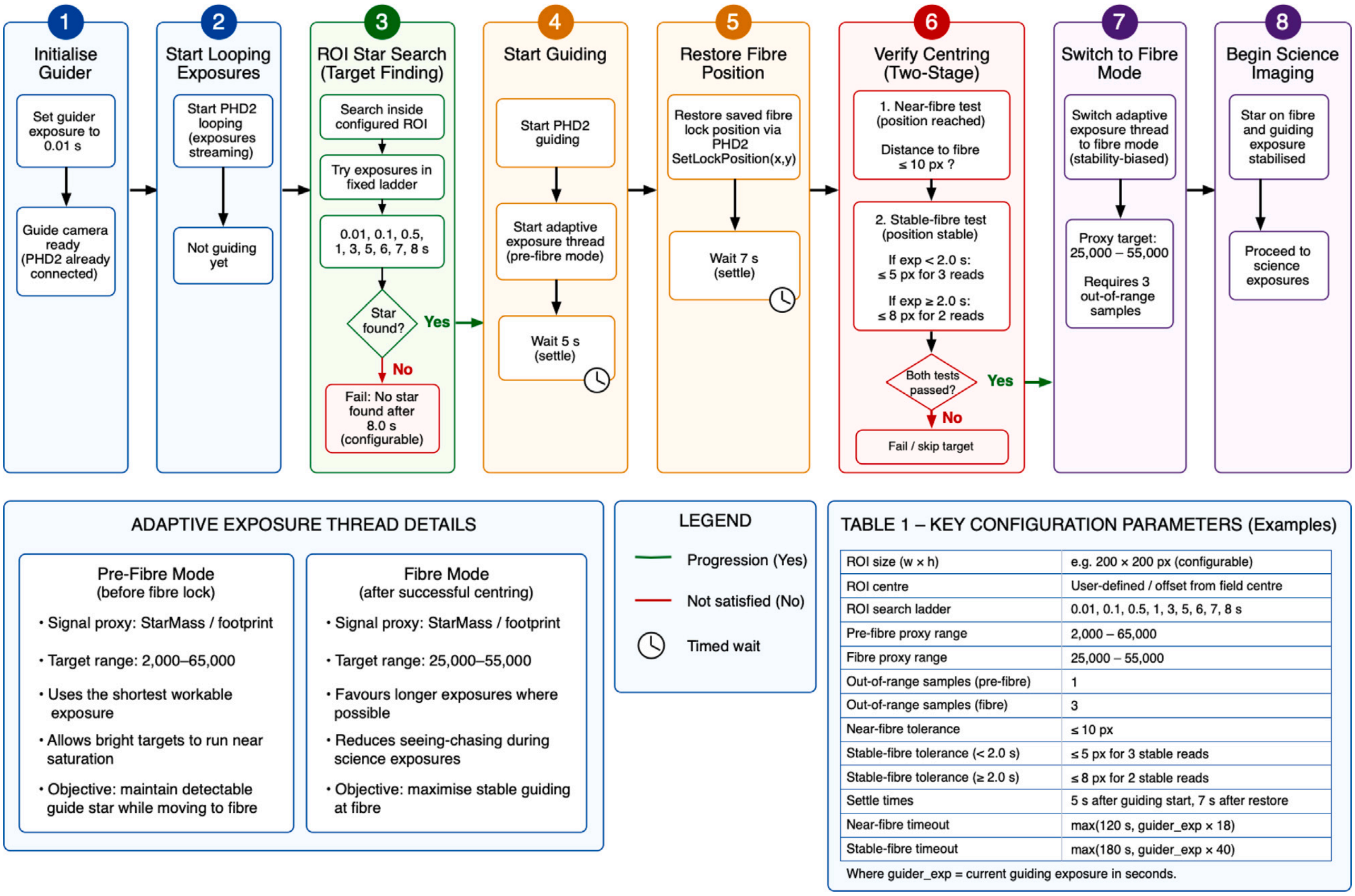

**Fig. 3.** Automated acquisition and guiding workflow.

The same exposure routine is used for both calibration and target imaging. In `gui_sequencer.py`, the `run_calibration()` function switches the calibration lamps through helpers in `utils.py`, waits briefly for lamp stabilisation, and then calls `expose_and_save()` for each tungsten and thorium frame. The `run_target()` routine ensures the lamps are off before science imaging begins and then calls the same exposure function for the requested number of target frames. In automated operation, `batch_runner.py` injects the target name, exposure time and frame count into the Sequencer tab and then calls these Sequencer routines rather than duplicating the camera-control code.

This design means that Spectro Capture does not simply trigger exposures; it controls the full path from camera exposure to reduction-ready FITS file. The same exposure and FITS writing pathway handles manual target runs, calibration blocks, added exposures and unattended batch imaging without the need for additional camera-control and acquisition software.

### 5.4. File and directory handling

The file-handling logic is matched to the author's reduction workflow but can be modified for other installations. Folder creation is centralised in `utils.py`, principally through the `nightly_folder()` and `next_run_folder()` helper functions. These create the specified structure within a spectroscopy root folder, using target specific folders, date coded 'night' folders and run folders where required.

The observing night folder uses an astronomical night convention rather than a simple calendar date boundary. Files acquired after midnight but before local noon are assigned to the previous evening's observing session, preventing a single night's data from being split across two calendar dates. Target names are also converted to file safe forms for directory and filename use, while preserving the original object name for SIMBAD resolution and FITS headers.

Additional run folders for a previously imaged target on the same night are also handled automatically. If a calibration or science run already exists, `next_run_folder()` creates sequentially numbered run folders rather than overwriting or mixing repeated runs. This allows for automated high cadence monitoring over several independent runs on a single target on the same night.

The default file handling in Spectro Capture keeps the data layout deterministic across manual, batch and scheduler-driven observing. Each run produces predictable folder locations and FITS filename sequences, reducing the amount of manual sorting required before data reduction.

### 5.5. Calibration frame acquisition

Calibration frame acquisition is implemented around the requirements of the Shelyak eShel spectrograph. The system acquires tungsten flat field frames and thorium wavelength calibration frames, with exposure settings entered on the Sequencer tab. Calibration acquisition is handled by the `run_calibration()` function, which reads these calibration settings and is used both for direct Sequencer calibration runs and for automated Batch Runner calibration blocks. Calibration save locations are created using the same folder helpers used elsewhere in the application.

During the calibration sequence, the routine switches on the tungsten lamp first, waits briefly for the lamp to stabilise, and then acquires the requested tungsten sequence using `expose_and_save()` in `sequencer_logic.py`. It then switches to the thorium lamp, again waits for stabilisation, and acquires the thorium sequence through the same camera control and FITS writing logic.

Lamp control is handled through the shared relay helper functions in `utils.py`, which call a local lamp control executable developed for the eShel lamp calibration module. The Sequencer also updates the shared application context to reflect the current tungsten and thorium lamp state, allowing the interface to display lamp state via visible status indicators. A final `lamp_off()` call is made at the end of the calibration routine so that the lamps are not left on after a completed or interrupted calibration block.

In an automated sequence, calibration frames are normally acquired after the telescope slew, but before guide acquisition begins. This is appropriate for the eShel configuration because the calibration lamps illuminate the spectrograph directly and are not compatible with simultaneous guide acquisition on the target. After the calibration block is complete, Batch Runner proceeds to PHD2 acquisition, fibre positioning and the science exposure sequence.

Batch Runner can be configured to control where calibration blocks are inserted into an observing sequence. A target can have its own pre-target calibration block, calibration can be skipped for selected targets, and scheduler mode can run a one-time calibration block before the first science target. There is also an optional shutdown calibration block after the batch completes. In each case, the calibration frames are still acquired through `run_calibration()` and saved through the same FITS generation path as science images.

This keeps calibration handling consistent with the rest of the system. Tungsten, thorium and science frames are acquired by the same camera backend, written with the same FITS logic, and stored using the same directory conventions. Batch Runner then controls when calibration blocks are inserted during automated operation, making calibration acquisition part of the same repeatable workflow as target acquisition and science exposure sequencing, rather than a separate observer-dependent task.

### *5.6. Batch operation*

The Batch tab has become the main operational interface in Spectro Capture. It provides a way to apply the previously described control logic to a prepared observing plan, allowing multiple targets to be processed automatically.

The Batch tab is implemented in `gui_batch.py` and presents a table containing a list of targets and observing parameters. Targets can be added or edited manually, disabled without being removed, or loaded from and saved to CSV files. This allows observing plans to be prepared outside the application and then loaded for execution at the telescope. In practice, these CSV files may contain large target lists and can be managed using external tools such as spreadsheet software. A target can be retained in the file but skipped during execution by setting its `Enabled` value to `no`. The same file format can include both sequential targets and timed targets, linking the Batch tab workflow with the scheduler behaviour described below.

A shortened example of the CSV input format is shown below.

```
Target,Exposure,Frames,Calibration,Enabled,StartUT,ReferenceStar
HD 155506,600,8,no,no,19:40,
v507 Car,600,9,no,no,12:53,
CD-3011814,600,12,no,no,17:02,HD 131625
[SC96] Mis 50,600,9,no,yes,,
2 Cet,600,6,no,yes,,
```

When a batch is started, the table rows are passed to `batch_runner.py`, where they are checked and converted into target blocks. Disabled targets are skipped, while invalid rows are rejected before execution. In conventional batch mode, enabled targets are processed sequentially in the order shown in the table.

For each enabled target, Batch Runner coordinates the telescope slew, settle, calibration, guiding and exposure routines. Once the target sequence is complete, guiding is stopped and Batch Runner advances to the next enabled target.

The optional reference star fields allow an associated reference star to be inserted immediately after a completed science target. The reference star observation uses the same target-execution pathway but suppresses calibration frames, keeping the reference star tied to the relevant science target without requiring a separate manually started sequence.

Individual targets may also have a start time, allowing the batch to wait before beginning a target. If there is a gap of more than two minutes before the next timed target, the telescope is automatically parked during the idle period. The Batch tab also provides optional end-of-run actions, including telescope park, dome close, shutdown calibration and camera warm up.

Throughout the run, Batch Runner monitors shared stop flag and dome safety checks. If a stop is requested, or if an unsafe dome condition is detected, the current operation is interrupted and the batch is not allowed to proceed blindly to the next target. At the end of execution, a batch summary is written to the log, listing completed science targets, reference targets and failed or skipped entries.

In this way, Batch Runner acts as the operational layer above the individual control modules. It does not replace the telescope, camera, guiding, calibration or file-handling routines, but coordinates them into a repeatable observing process. This is why it has become the de-facto mode of use: the same unattended pathway can be used for both short observing runs and full multi-target nights.

### *5.7. Scheduler logic*

The Scheduler module adds an additional selection layer to Batch Runner. It is enabled in the Batch tab via the *Smart HA Scheduling* option, together with user defined hour angle limits, an optional start time and optional calibration actions. These settings are passed to `batch_runner.py` through the shared application context.

The scheduling logic is deliberately separated from hardware control. The `scheduler.py` module decides which target is eligible to start next, while `batch_runner.py` remains responsible for executing the selected target through the normal batch pathway.

In conventional batch mode, enabled targets are run in the order shown in the Batch table, subject to any optional per-target start times. In Smart HA mode, Batch Runner builds its working target list when the batch is started, using the enabled rows and any start time fields in the Batch table. Timed targets are therefore mixed into the same observing list as ordinary HA-selected targets. During the run, `batch_runner.py` repeatedly re-evaluates the remaining targets while `scheduler.py` determines eligibility. A timed target is given priority once its start time has arrived, even if it is outside the configured hour-angle window. HA-selected targets are considered only when they fall within the configured hour-angle limits and can be completed without delaying an upcoming timed target. If no suitable target is available, Batch Runner parks the telescope, waits for a short interval, and then re-evaluates the batch list. This allows the software to select targets as they enter the eastern hour-angle limit, without requiring fixed start times for every target.

Smart HA Scheduling has proved useful for work with large target lists. In current testing, the feature has been used with a southern Be star observing list containing hundreds of targets. After each observing night, targets that have been successfully acquired are manually disabled, leaving the remaining enabled targets available for subsequent runs with minimal reconfiguration.

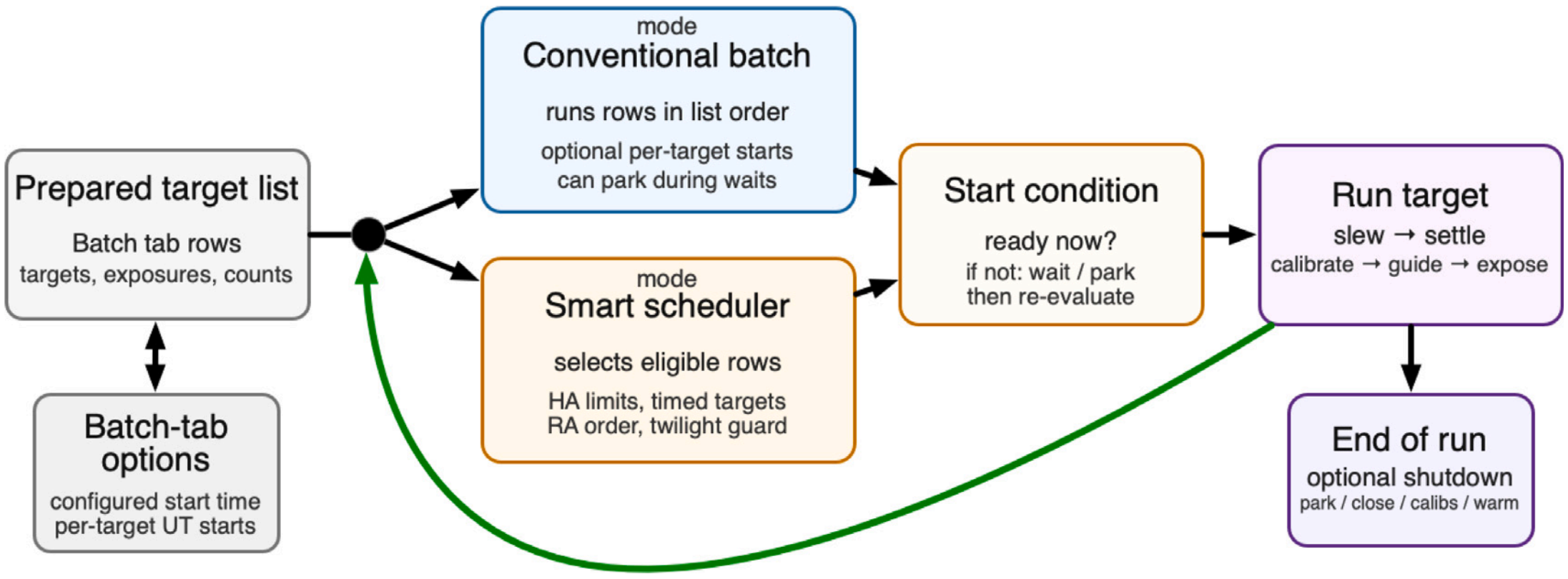

**Fig. 4.** The Batch Runner workflow.

A twilight guard provides the end-of-night boundary for scheduled operation. In Smart HA mode, `batch_runner.py` calculates a latest science-start time based on nautical dawn, with a margin before twilight.[4] New opportunistic targets are not started after this limit, although a target already in progress is allowed to finish.

Calibration options are handled by Batch Runner rather than by `scheduler.py`. The scheduler determines target eligibility and ordering, while the existing sequencer and calibration routines perform the lamp and exposure work.

### 5.8. Status indicators and logging

Status indicators on the graphical interface provide a compact view of the main hardware and observing states. The shared application context in `app_context.py` is used to update indicators for camera connection, telescope connection, dome state, guide state, lamp state, camera temperature and related run status. These indicators are not a separate control layer; they mirror state reported by the underlying camera, telescope, dome and lamp control routines so that the observer can see at a glance whether the system is ready, active or in an unsafe state.

Logging is also handled through the shared application context. General status messages are sent through `context.log()`, which writes to the Status tab and appends the same messages to a nightly disk log. Batch Runner also writes a summary at the end of each observing run log, listing completed science targets, reference targets and failed events.

A separate guide log records detailed PHD2, ROI acquisition, adaptive exposure and fibre settle messages on the Guide tab. During Batch Runner operation these guide messages are also written to a separate timestamped guide log file for later review.

## 6. Operational results and behaviour

Spectro Capture has been in continuous operation at the author's observatory since December 2025. The statistics in this section are derived from log files covering the period from 1 January onwards. During this period, observing was carried out exclusively with Spectro Capture utilising Batch Runner in Smart HA mode with some timed targets. Spectroscopic targets observed during this period included:

- an extended list of southern Be stars monitored as part of the BeSS project (BeSS Database, 2026)
- a small number of targets from the ARAS Eruptive Stars database (ARAS Spectral Database, 2026)
- several faint targets observed in collaboration with other researchers

Site conditions and instrument limitations restrict observable targets to magnitude 10 or brighter with typical total exposure integration times of 1–2 h depending on target brightness. A standard 600 s subexposure for targets fainter than magnitude 4.5 is commonly used, with a pre-determined sliding scale for bright targets that saturate at or near 600 s. At best, this achieved 7 targets per night over the late summer months.

The southern Be star list contains 497 targets, of which 200 have been successfully acquired in Smart HA mode. Timed starts were successfully incorporated into Smart HA mode for the additional project stars so that the calculated mid sequence corresponded with meridian transit, improving observing conditions where possible for these targets.

### 6.1. Log analysis

The nightly disk logs are the source for the completion statistics presented in Table 1. Early evening tests and manual runs have been excluded from the analysis in order to present an accurate picture of system performance in a completely unattended mode of operation. In most cases, these 'all night' runs finish just before dawn, with the Smart HA mode of operation automatically filling the schedule with Be star observations around any timed targets belonging to specific projects.

Between 1 January and 11 May 2026, Spectro Capture attempted 345 science targets and successfully completed 339 within these all night batch runs.

**Table 1**
Target completion statistics for batch runs between 1 January and 11 May 2026.

| Metric | Value |
|---|---|
| Number of batch runs | 74 |
| Science targets attempted | 345 |
| Science targets completed | 339 |
| Science target completion rate | 98.3% |

[4] The current implementation includes site-specific observatory coordinates used for sidereal-time, hour-angle and nautical-twilight calculations. These values are defined in `utils.py` and are used by the scheduling and twilight-guard logic. Use at another site would require these coordinates to be reviewed and updated.

In all cases, completed data were subsequently reduced to publishable spectral profiles and submitted to the relevant database or collaborators. The 98.3% completion rate therefore corresponds not merely to software sequence completion, but to successful acquisition of scientifically usable spectra.

The six science targets that did not reach the completed target sequence are listed in Table 2. Although extremely few, these cases do provide a useful check of target failure logging.

**Table 2**
Targets attempted during nightly batch runs that did not reach completion.

| Date | Target | Reason |
|---|---|---|
| 2026-01-21 | V518 Car | Fibre-restore failure |
| 2026-02-03 | HD 75658 | Fibre-restore failure |
| 2026-03-11 | V801 Cen | SIMBAD resolution failure |
| 2026-04-05 | V1008 Cen | Weather-safety interruption |
| 2026-05-03 | LV Mus | Fibre-restore failure |
| 2026-05-08 | mu Cen | Fibre-restore failure |

Review of the six non-completed targets indicates that four were associated with guide acquisition or fibre-restore failure, one with a rare SIMBAD access failure that prevented pre-slew coordinate resolution and one with a confirmed weather safety interruption which occurred after the Boltwood Cloud Sensor had been installed. It is therefore treated here as a successful interruption of an unsafe batch condition, not as part of a weather loss analysis.

Dedicated guide logs were introduced on 28 January 2026, allowing target acquisition and fibre settle behaviour to be analysed separately from the broader application logs. For this period, the main and guide logs were reconciled and, as expected, science targets that reached completion also had corresponding stable fibre acquisition recorded in the guide log.

### 6.2. Guiding performance

For completed science targets, the full automated acquisition overhead from PHD2 looping to the start of the first science exposure had an average duration of 62 s, with 90% of cases completed within 92 s. This indicates that target acquisition, fibre restoration and guide stabilisation typically add about one minute of overhead per science target. It is worth noting that fibre restoration is highly dependent on the distance between the ROI centre and the fibre position, which in the case of the current instrument geometry is 2–3 arc minutes.

**Table 3**
Acquisition and guide-settle timing for the 268 completed science targets in the dedicated guide-log period, 28 January–11 May 2026. The timing values characterise the software controlled acquisition and guide settle workflow, and should not be interpreted as a controlled comparison with manual target acquisition.

| Metric | Average | 90% completed within |
|---|---|---|
| PHD2 looping start → star found | 12 s | 15 s |
| Guiding started → fibre restoration complete | 46 s | 72 s |
| PHD2 looping start → first science exposure started | 62 s | 92 s |
| Average guide exposure pre-fibre mode | 0.5 s | 1.8 s |
| Average guide exposure on fibre | 6.9 s | 7.0 s |

## 7. Discussion

Within a relatively short deployment period, unattended nightly operation has become the normal observing mode, with all completed target sequences producing usable spectra. The significance of this cannot be overstated. The move from supervised single-target observing to unattended multi-target spectroscopy has significantly changed how the observatory is used. In routine use, automated acquisition appeared to be faster than manual target acquisition, although this was not measured as a controlled comparison. The relevant measure of efficiency in this study is not the acquisition time for a single target, but whether a prepared observing programme can continue reliably without repeated intervention.

The CSV-to-Batch Runner process has produced a reusable observing programme that can be revised between runs rather than rebuilt each night. By automating an already stable hardware system, Spectro Capture has reduced nightly setup to a short pre-twilight task, with the remainder of the run proceeding without further intervention.

Smart HA scheduling extends this concept by reducing the need to assign fixed times to every target and then continually revise those times over the course of an observing season. Instead, hour angle becomes the main criterion for deciding when an untimed target should be observed. Individual targets are processed when they become geometrically favourable, rather than being forced into a rigid timetable. Timed targets can still be included alongside opportunistic target selection from the enabled target list, providing even greater flexibility.

### 7.1. Adaptive acquisition and guiding

The acquisition and guiding results suggest that the most important design decision was not simply automating PHD2 but allowing guiding behaviour to adapt to changing target brightness and sky conditions. The two-stage adaptive exposure logic enables Batch Runner to handle targets within the observatory's practical magnitude limit with remarkable reliability.

### 7.2. Repeatability and local instrument configuration

The results also show that success depends on repeatability. Spectro Capture is not a general solution that can compensate for unstable hardware or changing optical geometry. It works because the telescope, mount, guide camera and spectrograph behave as a stable and calibrated system.

This is considered a strength not a constraint. The software can make useful assumptions because the instrument *is* stable. That stability allows automated target placement to be treated as a predictable process rather than a new problem that needs to be solved for every target. Even without automation, spectroscopic systems require a higher level of instrument stability than conventional astronomical imaging.

This is an important conclusion. The software does not replace the need for a stable installation. It makes use of one. For small-observatory spectroscopy, the more realistic path to automation is a software layer built around a stable and calibrated instrument.

### 7.3. Limitations and future development

As discussed, the present implementation remains specific to the author's system with operational assumptions that reflect the equipment and workflow at Shenton Park Observatory. Another observatory would need to establish its own relationship between slew position, guide-camera field and fibre position. These values are configurable in the scripts and software interface but still require testing and refinement for each installation. Spectro Capture is therefore transferable in principle but not necessarily plug-and-play.

The practical magnitude limit is real and has yet to be fully explored. The observing programme was deliberately weighted towards a practical site limit of magnitude 10 or brighter targets. Fainter targets are difficult not only because they require longer science exposures, but because in this instance they must also be automatically acquired and guided on. The few guide related failures would seem to mark the edge of present operations and future improvements might focus on behaviour at that boundary, such as longer faint-target patience and more flexible settle criteria.

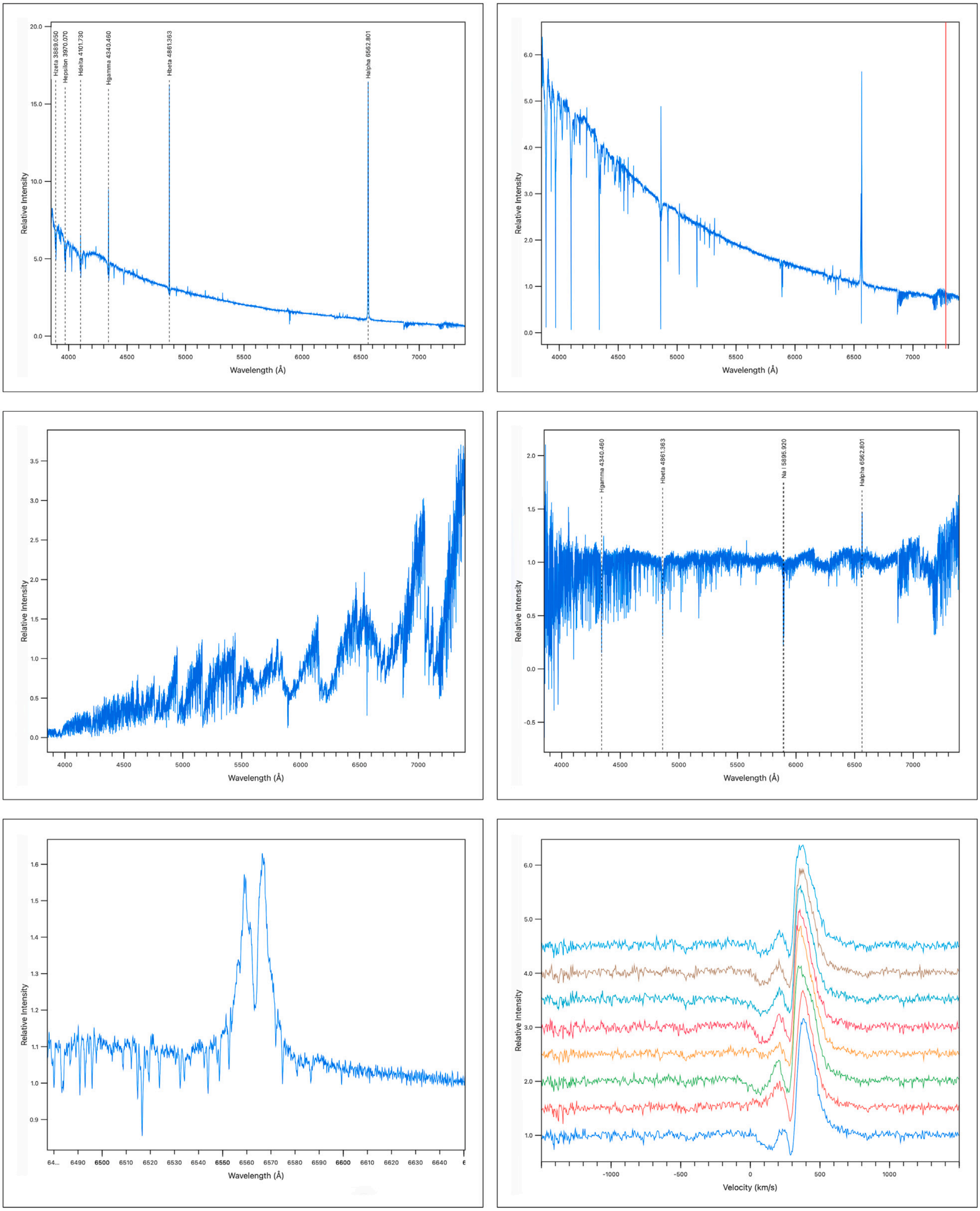

**Fig. 5.** Representative spectra acquired with Spectro Capture during unattended observing runs. The examples include emission-line Be stars and symbiotic stars, illustrating the range of targets observed using the automated workflow. The spectra were reduced using ISIS (Buil, 2026) at a resolving power of approximately $R \approx 13{,}000$. The final plots were prepared using the author's Spectro Plot software (Luckas, 2026). These and other spectra acquired by the author using this system are available through the BeSS database (BeSS Database, 2026) and the ARAS spectral database (ARAS Spectral Database, 2026).

Weather handling is intentionally conservative. The system can stop safely when unsafe conditions are detected, but it does not yet treat weather interruptions as recoverable events. A future version could pause, secure the observatory, monitor conditions and resume the remaining observing plan if safe conditions return. Because the Boltwood Cloud Sensor was introduced part way through the analysed period, the logs do not provide a uniform weather loss record across all runs. Weather interruption and recovery can therefore be evaluated more fully in future work.

Live SIMBAD dependence is another operational weakness. It is convenient during normal use, but it creates a potential point of failure for unattended operation. Some level of local coordinate storage or pre-run caching would make the system less vulnerable to network or service interruptions.

There is also considerable scope to clean up the now redundant Auto Capture mode, improve batch editing and enhance logging. The stop and abort pathways would also benefit from further consolidation, so that interrupted operations are handled more uniformly across the different observing modes. Further automation of batch lists, such as automatically disabling completed targets, could streamline routine use in a survey-style workflow. Similar workflow refinements could be made to reference star and calibration handling, including more flexible reference star acquisition, calibration-block scheduling, and a more intuitive way of specifying calibration-frame requirements. Smart HA scheduling could also be extended with additional selection constraints, such as declination, magnitude, priority or target class and even moon avoidance, so that very large target lists can be used and filtered more intelligently.

## 8. Conclusion

The development of Spectro Capture set out to explore whether existing general purpose observing software could be replaced and enhanced with direct Python control. It has since evolved into an end-to-end system for automated target selection, acquisition and data capture.

Until recently, fully automated small-observatory spectroscopy was often treated as difficult enough to be effectively out of reach. The results presented here suggest that this assumption is beginning to change. Spectro Capture is one example of this shift, but it is not an isolated one: other advanced amateur and small-observatory operators are also developing similar approaches to automated spectroscopic acquisition. Together, these efforts suggest that unattended spectroscopy is moving from an aspiration to a practical operating mode for a variety of suitably stable instruments.

## Declaration of competing interest

The author declares that they have no known competing financial interests or personal relationships that could have appeared to influence the work reported in this paper.

## Appendix A. Code availability

The Spectro Capture source code is publicly available on GitHub. Licensing and attribution information are provided in the repository README and accompanying documentation. The code is released under a non-commercial source license permitting personal, educational and research use, with redistribution permitted for non-commercial purposes subject to attribution and retention of the license terms.

The repository can be accessed at: https://github.com/pluckas/spectro-capture

The software is provided as source code rather than as a packaged installer. Use of the application therefore requires a working Python environment and sufficient familiarity with Python to install dependencies, edit local configuration values and run the relevant scripts. Required Python packages and other dependencies are listed in the repository documentation.

At the author's observatory, Spectro Capture is currently operated on Windows 11 (version 25H2) using Python 3.11.4. Other Windows and Python versions may work, but have not been tested.

## Appendix B. Graphical user interface

### B.1. Descriptions

The Spectro Capture interface is organised into task-based tabs, arranged from left to right across the main window.

| Tab | Description |
|---|---|
| Sequencer | Target-name entry and manual/auto mode selection; calibration controls for tungsten and thorium frames; science exposure/count controls; legacy Auto Capture controls; single-image capture and Stop. |
| Batch | Main unattended observing interface. Provides the target table, CSV load/save controls, Run/Stop controls, Smart HA Scheduling options and end-of-run options including telescope park, dome close, camera warm-up and shutdown calibration. |
| Guide | PHD2 connection and guide controls. Stores the fibre lock-position X/Y reference coordinates, provides adjustment buttons and displays live guide messages. |
| Status | Live system log, mirrored to the nightly status log written to disk. |
| FITS Viewer | Quick inspection of the most recent FITS image using automatic stretch and maximum ADU readout. |
| Targets | Optional manual target lookup and telescope slew controls. A favourites table stores commonly used targets and coordinates. |
| Tools | Manual bias/dark utilities and manual lamp control for the eShel tungsten and thorium lamps via `lamp_on()` and `lamp_off()` in `utils.py`, which call the external `lamp_control.exe` relay-control application. |
| Dome | Dome connection and manual control through `gui_dome.py` and `dome_backend.py`, interfacing with the ASCOM dome driver. |
| Setup | Hardware connection and configuration for camera and telescope, ROI centre/size fields and image-root path selection. |

### B.2. Screenshots

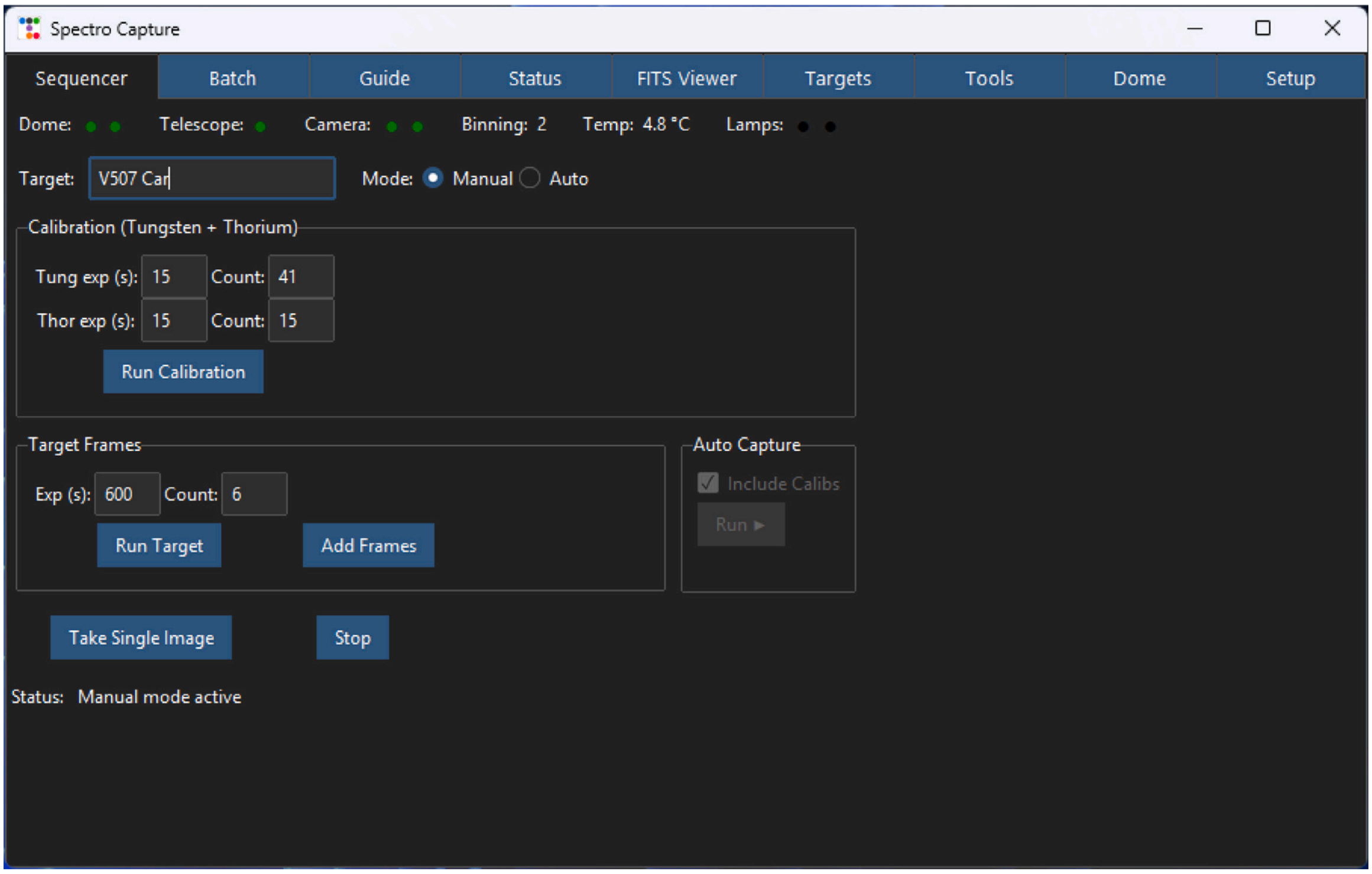

**Fig. 6.** The sequencer tab contains calibration controls and legacy single target capture modes.

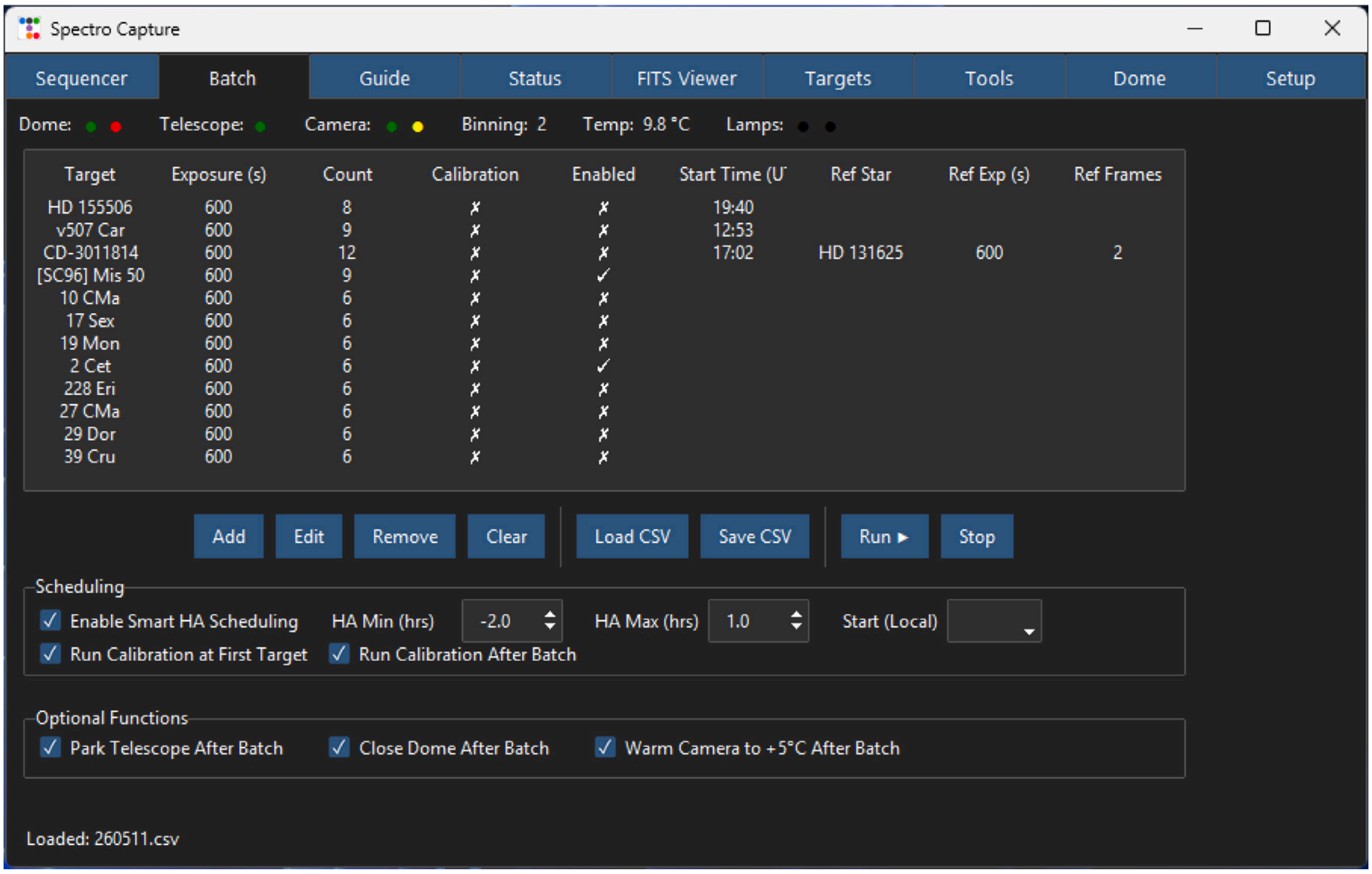

**Fig. 7.** The Batch Runner tab: The main observing interface for timed and scheduled targets.

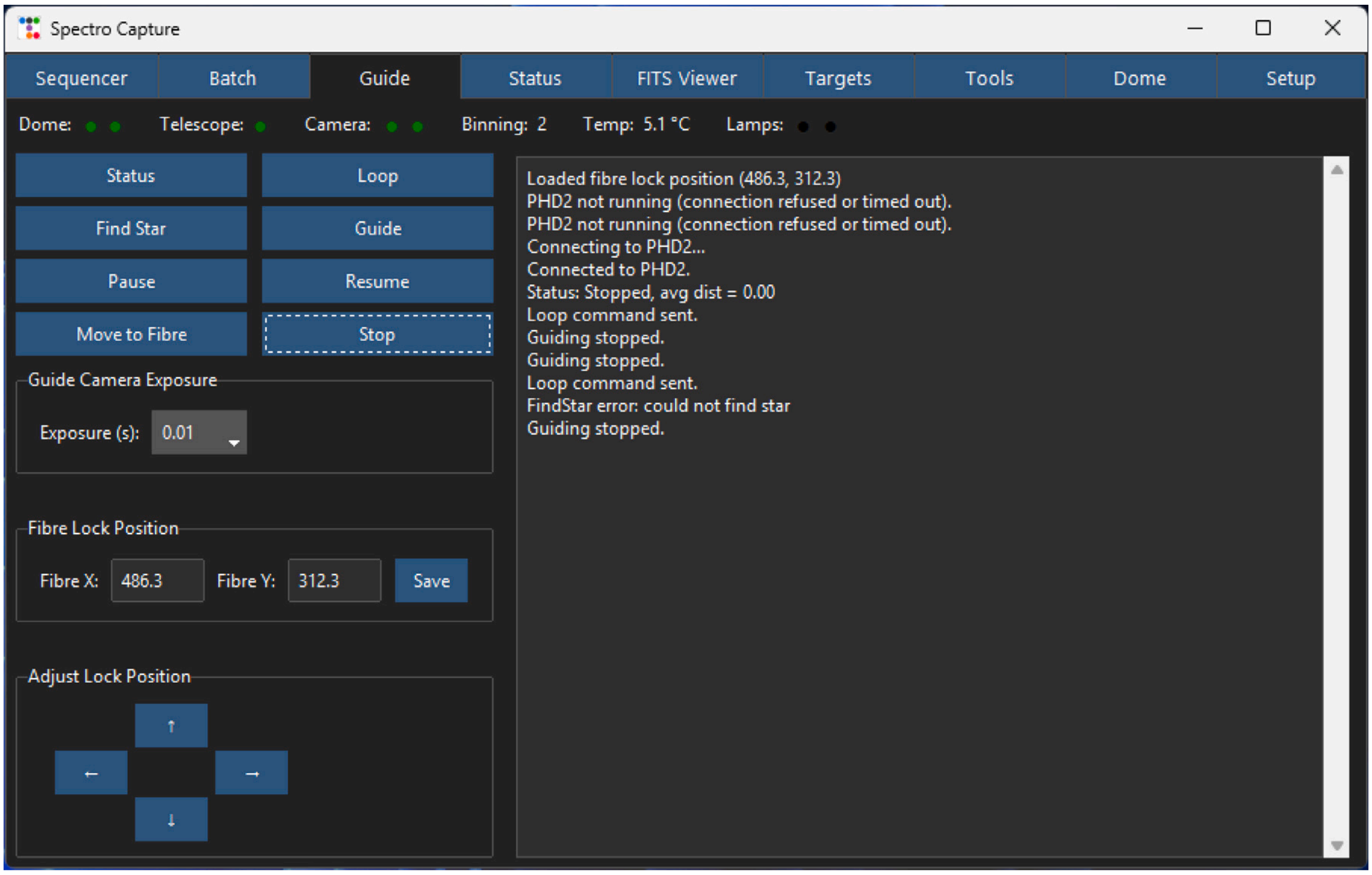

**Fig. 8.** The guide tab: Access to PHD2 controls, configuration and live guide information.

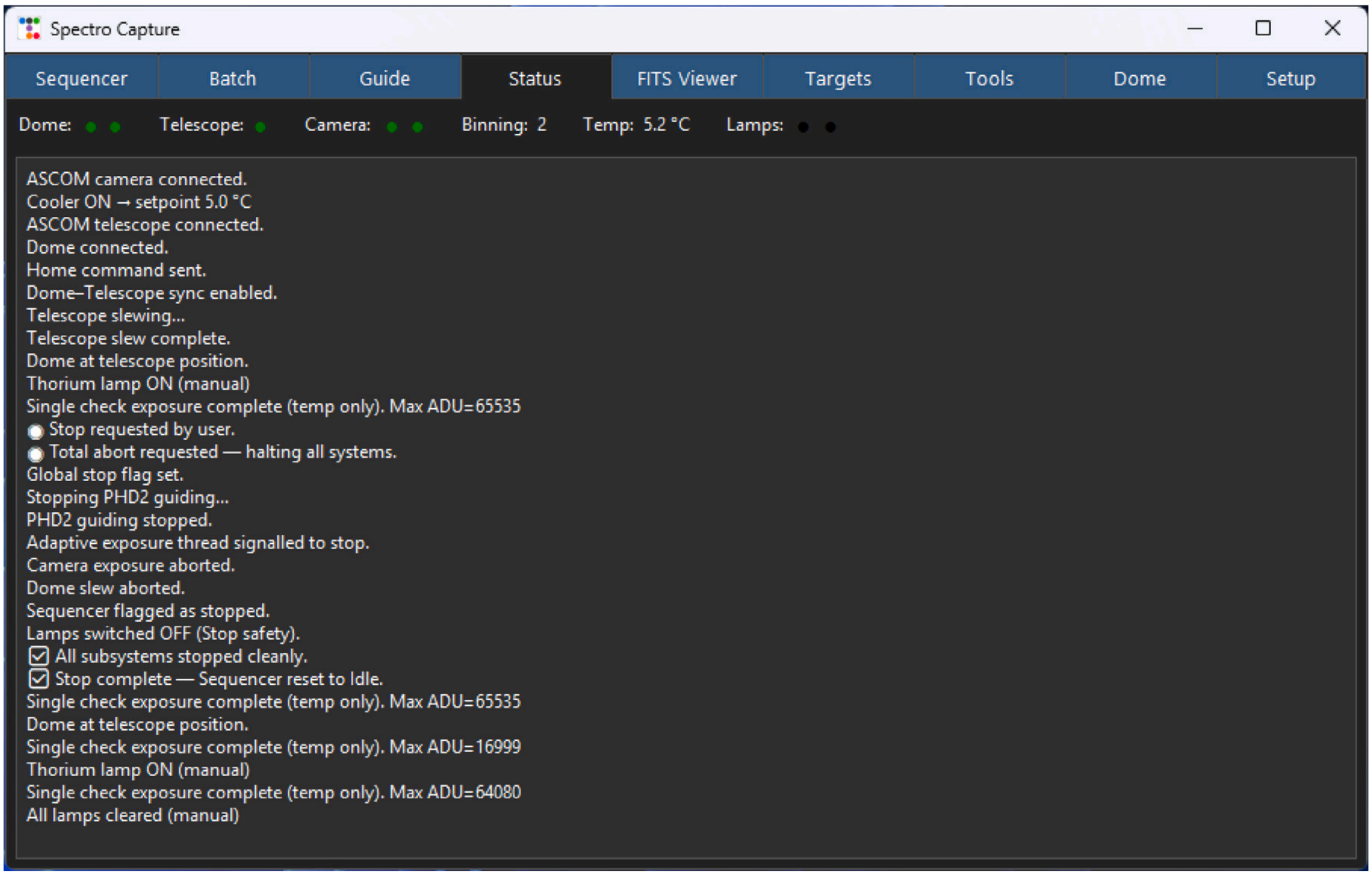

**Fig. 9.** The status tab provides live status information which is saved to a nightly disk log.

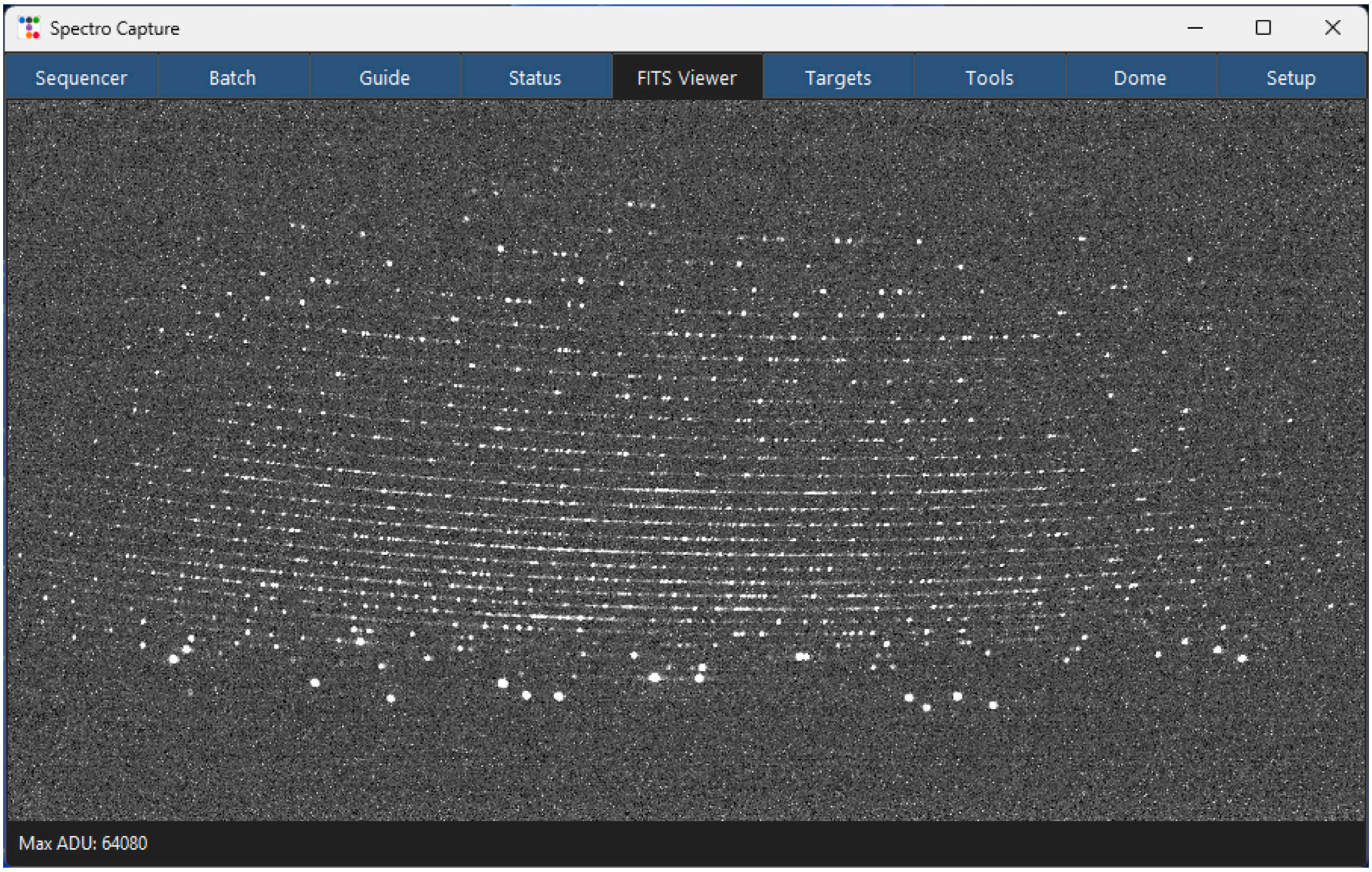

**Fig. 10.** The live FITS viewer with maximum ADU readout.

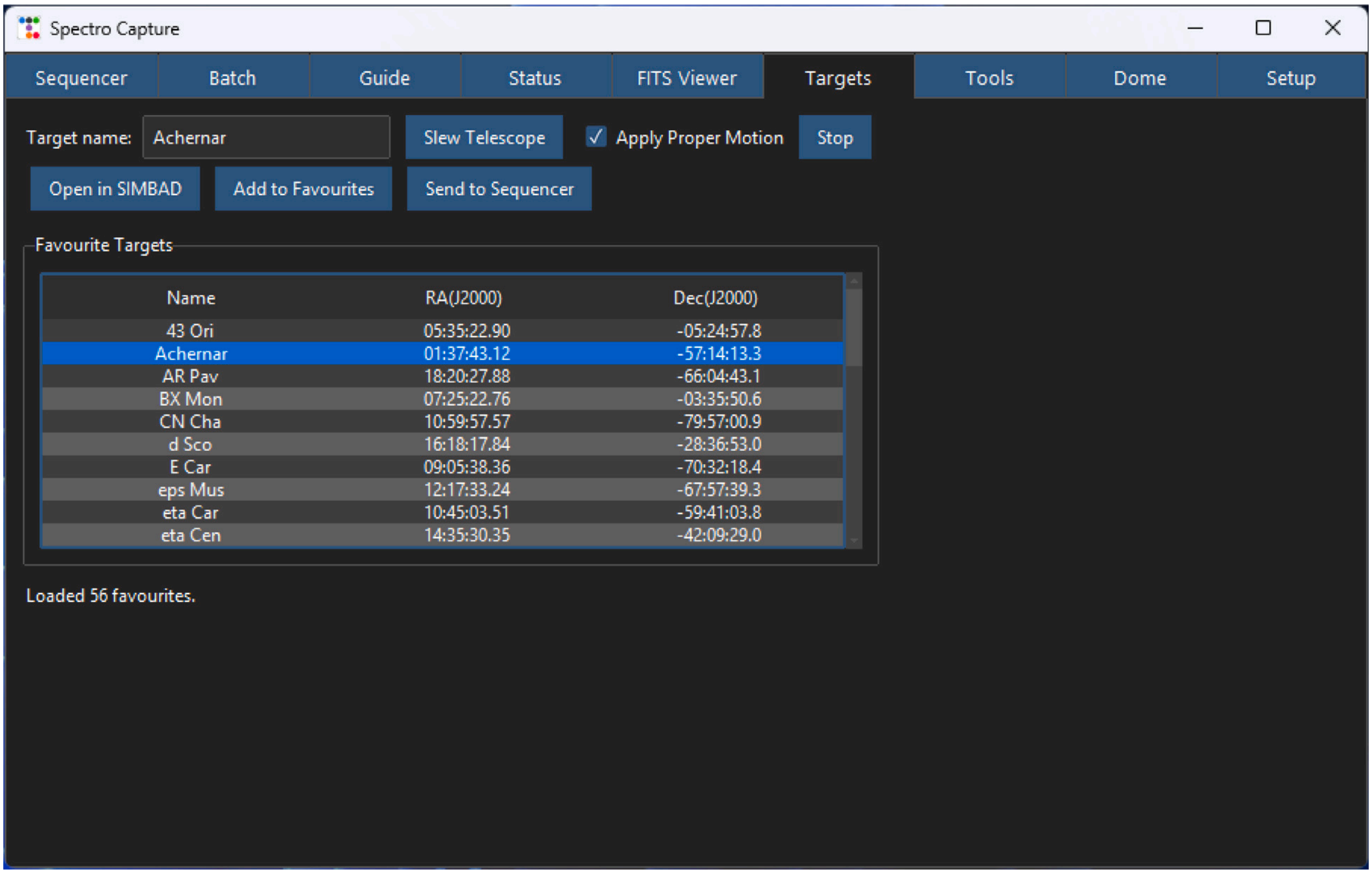

**Fig. 11.** The targets tab provides optional target lookup and control.

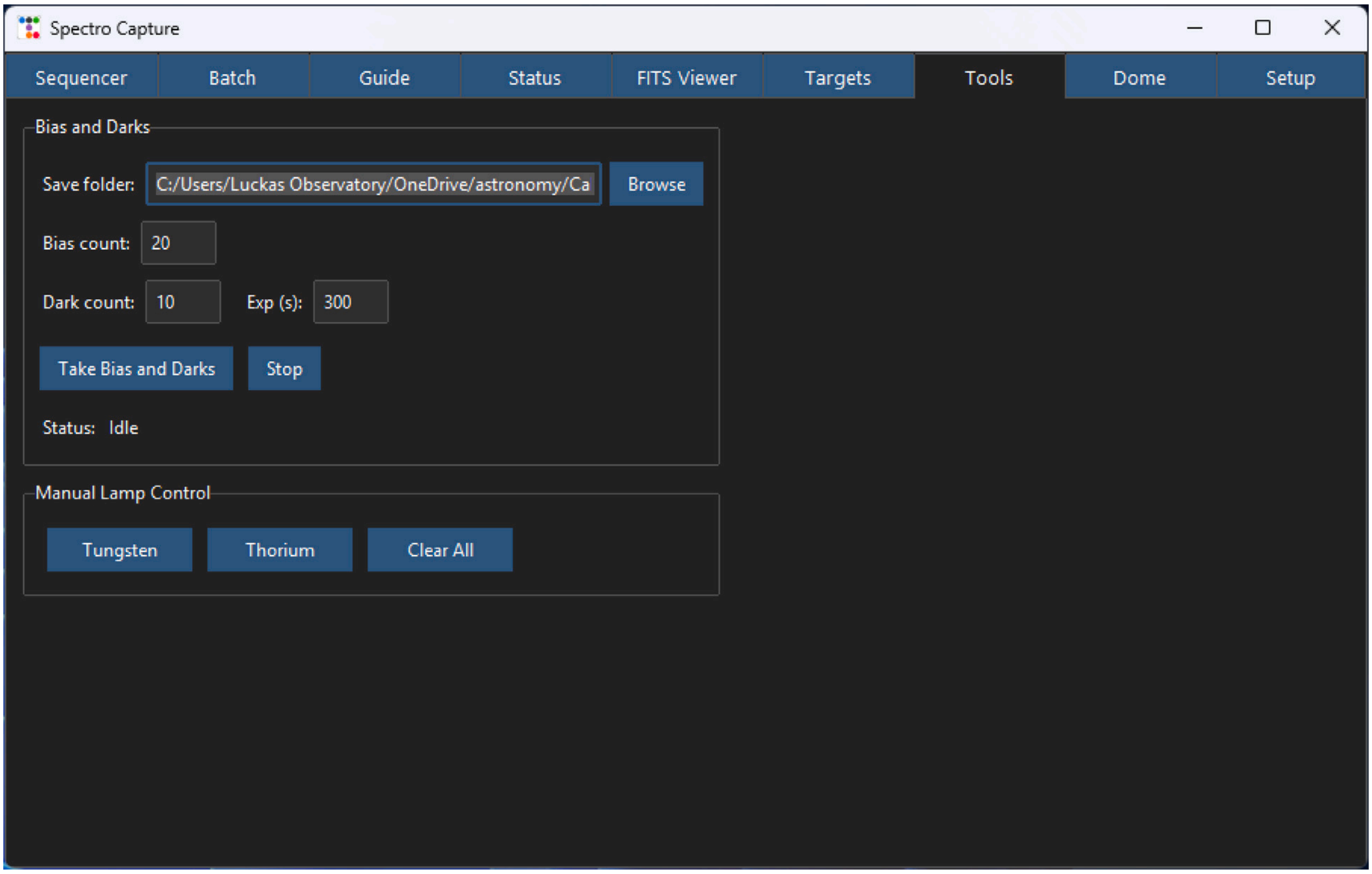

**Fig. 12.** The tools tab: Access to bias and dark frame acquisition and manual lamp control.

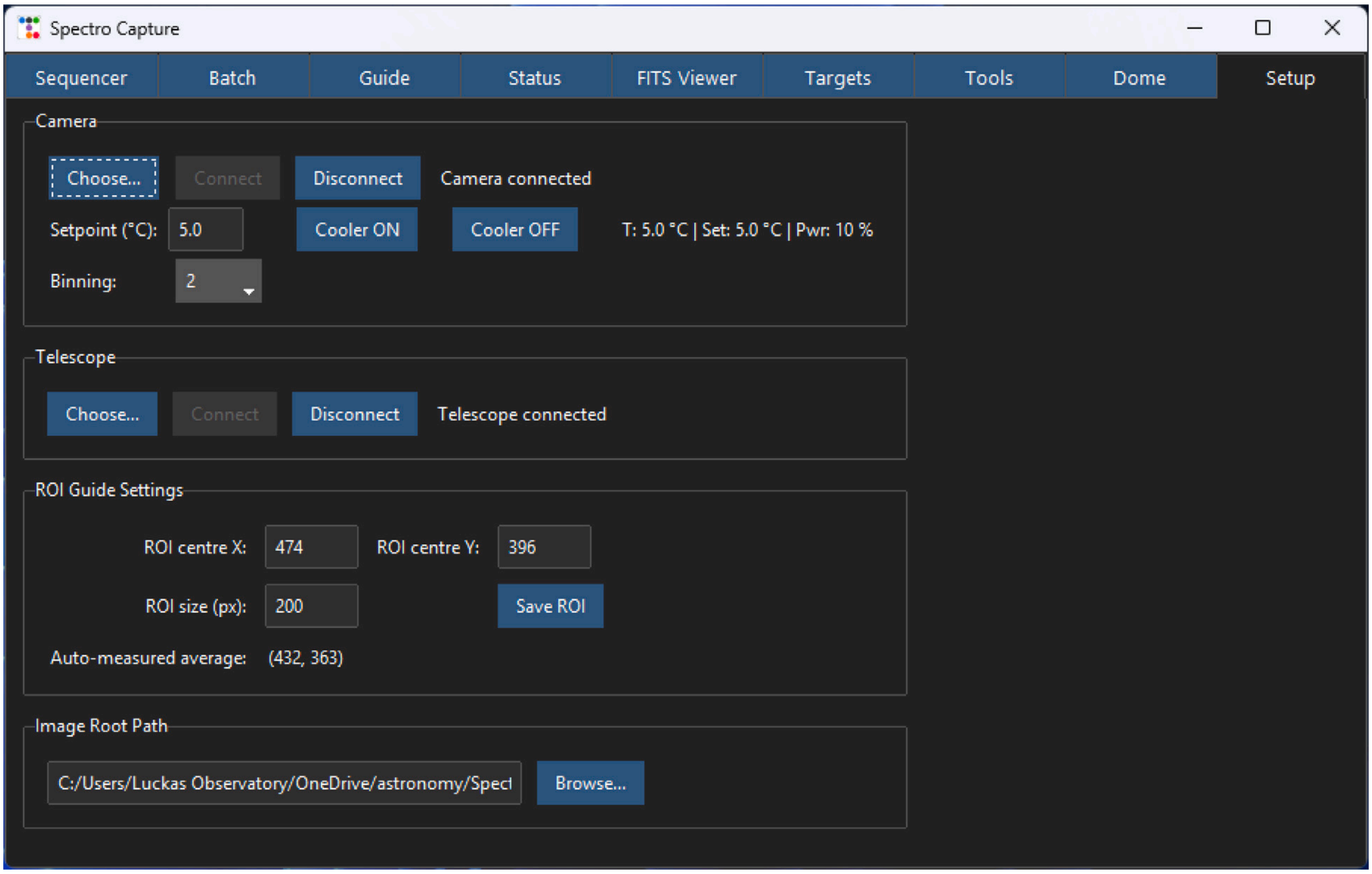

**Fig. 13.** Hardware setup and system configuration.

## Appendix C. Example automation log excerpts

### C.1. Example main log excerpt

The following excerpt shows a representative successful unattended target block. Long local file paths have been shortened for readability.

```
00:17:08 [3/207] Starting target 'HD 135160'
00:17:08 Sequencer updated: HD 135160, 600.0s x 6 frames.
00:17:08 Initiating PWI4 slew for target: HD 135160
00:17:08 Resolving 'HD 135160' via SIMBAD (with proper motion)...
00:17:09 Resolved HD 135160: RA=15 16 36.68, Dec=-60 54 14.48
          (pmRA=-2.91, pmDec=-3.00 mas/yr)
00:17:09 Slewing to HD 135160 via PWI4 HTTP (apply_pm=True)...
00:17:09 Slew to HD 135160 complete.
00:17:09 Waiting for telescope and dome to settle before guiding...
00:17:10 Telescope slewing...
00:17:11 Still moving: telescope
00:17:14 Still moving: telescope
00:17:16 Telescope slew complete.
00:17:18 Still moving: dome
00:17:21 Dome at telescope position.
00:17:25 Telescope and dome settled.
00:17:25 Skipping calibration frames (unchecked).
00:17:25 Checking dome state before starting guider...
00:17:25 Dome is stationary -- safe to start guider.
00:18:29 Sequencer updated: HD 135160, 600.0s x 6 frames.
00:18:29 Running target sequence for HD 135160 ...
00:18:29 Copied calib from ...\V795Cen\260511\calib
         to ...\HD135160\260511\calib
00:18:32 Lamp OFF (safety before target run)
00:18:32 Starting exposure 1 of 6
00:18:32 Exposing HD 135160-1.fit for 600.0s...
00:28:34 Saved ...\HD135160\260511\HD135160-1.fit
00:28:34 Starting exposure 2 of 6
00:28:34 Exposing HD 135160-2.fit for 600.0s...
00:38:36 Saved ...\HD135160\260511\HD135160-2.fit
00:38:36 Starting exposure 3 of 6
00:38:36 Exposing HD 135160-3.fit for 600.0s...
00:48:38 Saved ...\HD135160\260511\HD135160-3.fit
00:48:38 Starting exposure 4 of 6
00:48:38 Exposing HD 135160-4.fit for 600.0s...
00:58:40 Saved ...\HD135160\260511\HD135160-4.fit
00:58:40 Starting exposure 5 of 6
00:58:40 Exposing HD 135160-5.fit for 600.0s...
01:08:42 Saved ...\HD135160\260511\HD135160-5.fit
01:08:42 Starting exposure 6 of 6
01:08:42 Exposing HD 135160-6.fit for 600.0s...
01:18:44 Saved ...\HD135160\260511\HD135160-6.fit
01:18:45 Target block finished.
01:18:45 Stopping PHD2 guiding...
01:18:46 Target 'HD 135160' complete.
```

### C.2. Example guiding log excerpt

The following excerpt shows the corresponding guide log for the same target. It illustrates ROI acquisition, fibre-lock restoration, fibre-settle confirmation and adaptive guide-exposure control.

```
00:17:25 ===== GUIDE START: HD 135160 =====

00:17:25 Setting guider exposure to 0.01 s before starting loop...
00:17:27 Starting looping exposures...
00:17:33 Attempting smart ROI star acquisition (multi-exposure test)
...
00:17:33 Trying exposure 0.01s in ROI [374, 296, 200, 200]...
00:17:35 Trial 0.01s failed: could not find star
00:17:35 Trying exposure 0.10s in ROI [374, 296, 200, 200]...
00:17:37 ROI-based star search issued.
00:17:39 Star found in ROI at 0.10s -- continuing. (x=414.5, y=380.0)
00:17:39 ROI star acquisition successful -- proceeding to guiding.
00:17:39 Guiding started.
00:17:39 Waiting 5 s for guider to stabilise before fibre lock
restore...
00:17:39 Adaptive exposure thread active (starting 0.10s)
00:17:44 Restoring fibre lock position via GuideTab method...
00:17:44 Fibre lock restore command issued successfully.
00:17:51 Waiting for star to reach fibre region (<=10px)...
00:18:20 Star reached fibre region.
00:18:20 Refining fibre lock (tolerance=5px)...
00:18:29 Guider stable on fibre (within fine tolerance).
00:18:29 Switching adaptive exposure to FIBRE mode
(LOCK_MIN/LOCK_MAX).
00:18:31 -> Increasing exposure to 0.20s
00:18:35 -> Increasing exposure to 0.50s
00:18:39 -> Increasing exposure to 1.00s
00:18:45 Exposure Locked (within optimal ADU range)
00:19:41 -> Increasing exposure to 1.50s
00:19:46 Exposure Locked (within optimal ADU range)
01:04:27 -> Decreasing exposure to 1.00s
01:04:29 Exposure Locked (within optimal ADU range)
01:06:09 -> Increasing exposure to 1.50s
01:06:12 Exposure Locked (within optimal ADU range)
01:17:42 -> Decreasing exposure to 1.00s
01:17:46 Exposure Locked (within optimal ADU range)
01:18:46 PHD2 guiding session ended.
```